\documentclass[letterpaper,10pt]{article} 

\usepackage{osameet3} 

\usepackage{amsmath,amssymb}
\usepackage[colorlinks=true,bookmarks=false,citecolor=blue,urlcolor=blue]{hyperref}

\usepackage{wrapfig} 

\definecolor{myDarkGreen}{rgb}{0.00000,0.48824,0.00000}

\DeclareMathAlphabet{\pazocal}{OMS}{zplm}{m}{n}
\SetMathAlphabet\pazocal{bold}{OMS}{zplm}{bx}{n}

\def\nch{N_{\text{ch}}}
\def\bchsp{\Delta f}

\def\calA{\pazocal{A}}

\def\emax{E^\bullet}

\usepackage{tikz}
    \usetikzlibrary{positioning}
    \usetikzlibrary{calc}
    \usetikzlibrary{shapes}
    \usetikzlibrary{patterns}
    
\usepackage{pgfplots}
    \pgfplotsset{compat=1.15}
    \usepgfplotslibrary{fillbetween}
    \usepackage{pgfplotstable}

\def\emax{E_{\max}}

\begin{document}

\title{Mitigating Nonlinear Interference by Limiting Energy Variations in Sphere Shaping}
\author{
    Yunus Can G\"{u}ltekin\textsuperscript{(1)},
    Alex~Alvarado\textsuperscript{(1)}, 
    Olga~Vassilieva\textsuperscript{(2)},
    Inwoong~Kim\textsuperscript{(2)}, \\
    Paparao Palacharla\textsuperscript{(2)}, 
    Chigo~M.~Okonkwo\textsuperscript{(1)}, 
    Frans~M.~J.~Willems\textsuperscript{(1)}
}
\address{\textsuperscript{(1)}Information and Communication Theory Lab, Eindhoven University of Technology, the Netherlands \\ 
\textsuperscript{(2)}Fujitsu Network Communications Inc., Richardson, 75082 TX, USA}
\email{y.c.g.gultekin@tue.nl}

\copyrightyear{2022}

\vspace{-0.1cm}

\begin{abstract}
Band-trellis enumerative sphere shaping is proposed to decrease the energy variations in channel input sequences.
Against sphere shaping, 0.74 dB SNR gain and up to 9\% increase in data rates are demonstrated for single-span systems.  
\end{abstract}





\vspace{-0.005cm}

\section{Motivation}
Nonlinear interference (NLI) generated during the propagation of an optical signal over the fiber can be modeled as time-varying inter-symbol interference~\cite{Secondini2013_AIRinNLWDMwArbitrary}.
Thus, the NLI experienced by a symbol transmitted at a given time instant depends on the adjacent symbols, creating memory in the channel. 
The presence of memory in the NL fiber channel implies that {\it good} input sequences having certain temporal structures could experience a lower NLI.
Motivated by this, temporal shaping was studied in~\cite{Dar2014_ShapingGainforFOchan,Yankov2017_TemporalPS_jlt,wu2021_arxiv_ListCCDM}. 
More recently, achievable rates of communication over the fiber channel were computed based on the selection and transmission of {\it good} sequences~\cite{Civelli2021_SequenceSelectionBound}.
According to~\cite{Geller2016_ShapingNLphaseNoise,Cho2021_ShapingLightwaves}, variations in the energy of the transmitted signal are important in the process of creating NLI.
Thus, it is expected that symbol sequences with limited energy variations would experience lower NLI.

\begin{figure*}[h]
    \centering
\resizebox{0.8\columnwidth}{!}{\includegraphics[]{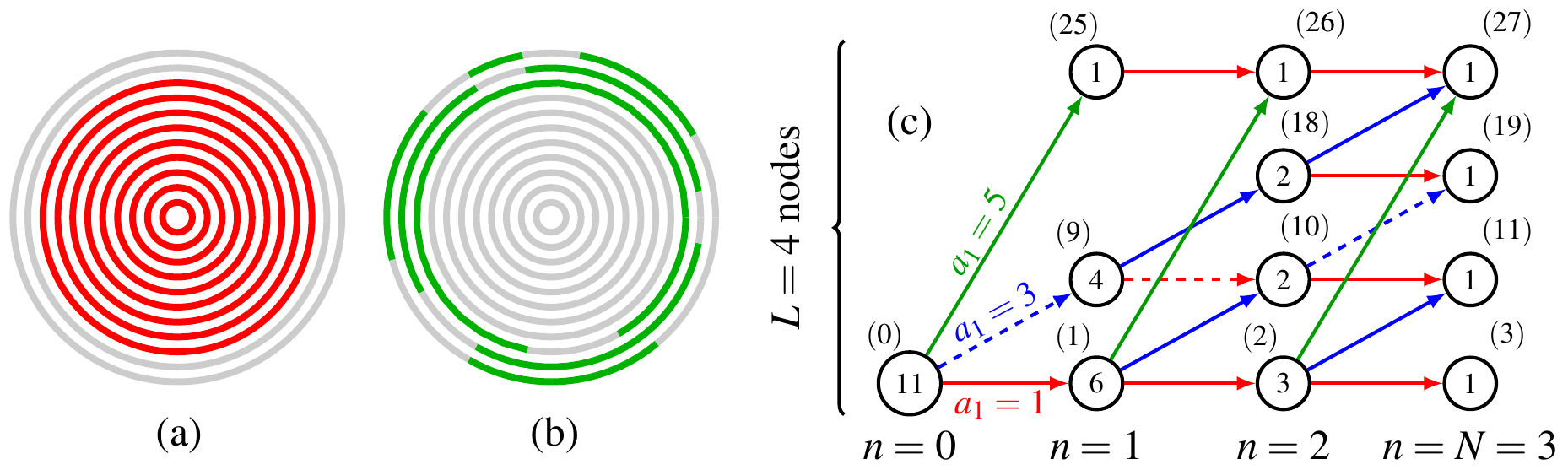}}
    \caption{(a) ESS. (b) B-ESS. (c) Enumerative trellis for $N=3$, $\calA=\{1, 3, 5\}$ and $\emax=27$.}
    \label{fig:esstrellisses}
\end{figure*}

In this work, we propose a novel algorithm for generating shaped input sequences with limited energy variations. 
We call this algorithm band-trellis enumerative sphere shaping (B-ESS), which is a generalization of ESS~\cite{GultekinHKW2019_ESSforShortWlessComm}. 
ESS operates based on a trellis that represents signal points within a sphere as shown in Fig.~\ref{fig:esstrellisses} (a).
We modify ESS such that only a {\it band}-like portion of the trellis is considered, in which only the sequences with small variations in energy are represented. 
These signal points are located on (but do not fill) a number of outermost shells of the sphere as in Fig.~\ref{fig:esstrellisses} (b).
Our main contributions are two: (i) we propose a new shaping algorithm (B-ESS), and (ii) we validate via end-to-end simulations that B-ESS experiences a reduced NLI for single-span systems.

\vspace{-0.7cm}

\section{Energy-band-limited Enumerative Sphere Shaping}
ESS assigns an index to the amplitude sequences $a^N \in \calA^N$ in an $N$-sphere, $\sum_i a_i^2 \leq \emax$, where $\calA$ is the alphabet~\cite{GultekinHKW2019_ESSforShortWlessComm}.
To this end, an amplitude trellis is created as shown in Fig.~\ref{fig:esstrellisses} (c), which is specified by $N$, $\calA$, and the number $L = (\emax-N)/8+1$ of nodes in the final column.
The nodes $(n,e)$ are labeled with the energy level $(e)$ that they represent, which is the accumulated energy of the sequences for their first $n$ amplitudes.
Branches that arrive at a node in $n^{\text{th}}$ column represent $a_n$.
Each sequence is represented by an $N$-branch path, starting from node $(0,0)$ and arriving at a node in the final column.
As an example, the dashed path in Fig.~\ref{fig:esstrellisses} (c) represents $(3, 1, 3)$ with energy $19$.
The numbers written inside each node $(n,e)$ give the number of ways to arrive at the final column, starting from node $(n,e)$.
Thus, the value inside $(0,0)$ is the number of sequences represented in the trellis, which is 11 for Fig.~\ref{fig:esstrellisses} (c).
Based on this trellis, ESS realizes a mapping from binary strings to amplitude sequences~\cite{GultekinHKW2019_ESSforShortWlessComm}.

\begin{figure*}[h]
\centering
\resizebox{\textwidth}{!}{\includegraphics[]{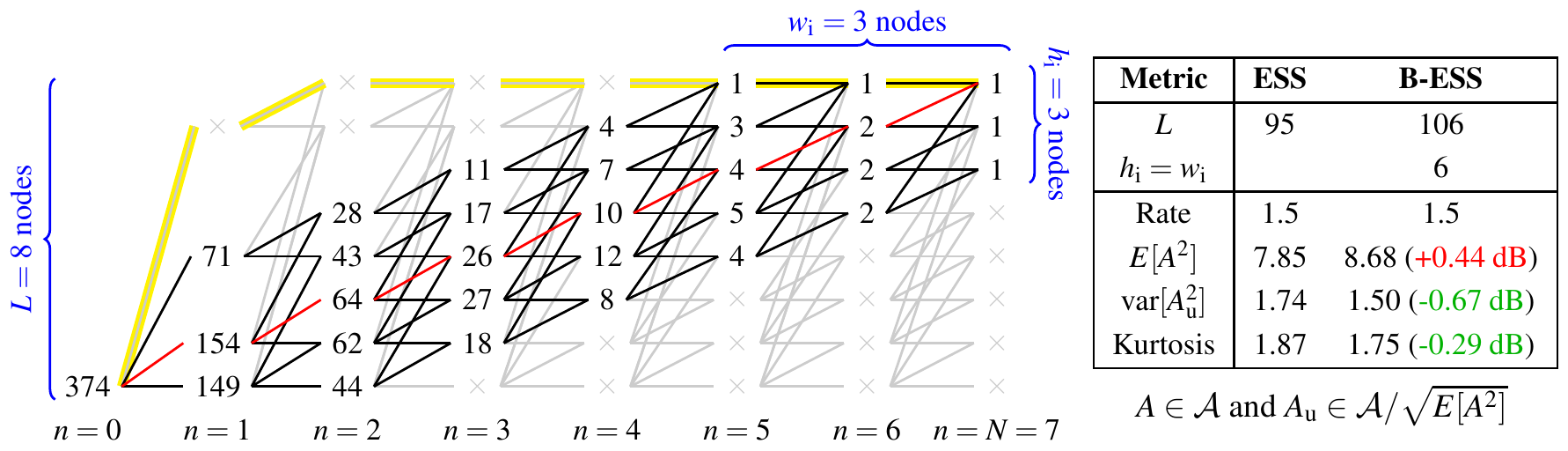}}
    \vspace{-0.5cm}
    \caption{Using $\calA=\{1, 3, 5, 7\}$: {\bf (Left)} Band-trellis for $\emax=63$. {\bf (Right)} ESS vs. B-ESS at $N=108$.}
    \label{fig:trellisses}
\end{figure*}

To discard sequences with large energy variations, we consider a band-like portion of the amplitude trellis as shown in Fig.~\ref{fig:trellisses} (left).
The motivation behind this is that the sequences represented in this band will have smaller energy variance $\text{var}[A^2]$ than the others. 
As an example, consider the sequence $(7,3,1,1,1,1,1)$ which is highlighted by yellow in Fig.~\ref{fig:trellisses} (left).
This sequence is in the complete trellis, but not in the band.
Also, consider the sequence $(3,3,3,3,3,3,3)$ which is drawn with red and belongs to the band.
The first sequence has an energy variance $\text{var}[A^2]=274.29$, while the second has no variations. 
Thus, we expect the sequences in the band to have a lower average energy variance than the sequences in the complete sphere.
This expectation agrees with the observation made in~\cite[Fig. 4]{wu2021_arxiv_EDI} on the relation between the trellis band-width and energy variations.
We specify the band-trellis with two values (in addition to $N, \calA, L$): 
the number of active nodes $h_{\text{i}}$ and $w_{\text{i}}$ in the top-right portion of the trellis.
Based on this band-trellis, ESS (now called B-ESS) again realizes a lexicographical mapping.

\vspace{-0.3cm}

In Fig.~\ref{fig:trellisses} (right), we compare ESS and B-ESS at a shaping rate of 1.5 bit/amplitude for $N=108$.
We see that B-ESS leads to a 0.44 dB degradation in average energy and thus, it is expected to perform worse than ESS in the linear regime.
However, the energy variance for B-ESS is 0.67 dB smaller than that of ESS, from which we expect gains in the nonlinear regime.
Finally, B-ESS also has a smaller kurtosis, which was recently demonstrated to improve the performance in the nonlinear regime~\cite{gultekin_kess_arxiv}.

\section{End-to-end Decoding Results}
We simulated optical transmission over a single span of fiber using split-step Fourier method with an attenuation of 0.2 dB/km, a dispersion parameter of 17 ps/nm/km, and a nonlinear parameter of 1.3 1/W/km, followed by an erbium-doped fiber amplifier with a noise figure of 5 dB. 
The transmitter generates a dual-polarized 50 Gbaud 64-QAM signal with a root-raised-cosine pulse with 10\% roll-off.
For the multi-channel scenario, the number of wavelength-division multiplexing (WDM) channels $\nch=5$, the channel spacing $\bchsp$ is either 55 or 75 GHz.
The 648-bit low-density parity-check codes of the IEEE 802.11 are used as the channel codes.
The target information rate is 8-bit/4D (400 Gbps) which is achieved using the rate-2/3 code for uniform signaling, and the rate-5/6 code for probabilistic amplitude shaping (PAS)~\cite{BochererSS2015_ProbAmpShap}.
For PAS, we realized kurtosis-limited ESS (K-ESS)~\cite{gultekin_kess_arxiv} at $N=108$, along with ESS and B-ESS which are realized using the parameters listed above for Fig.~\ref{fig:trellisses} (right).

\begin{figure}[h]
\centering
\resizebox{7cm}{!}{\includegraphics[]{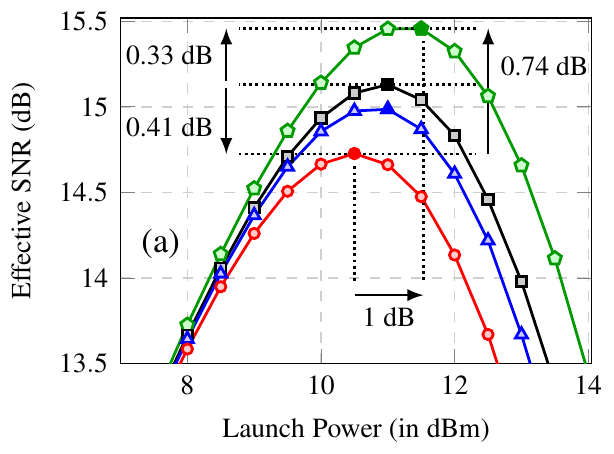}}
\resizebox{7cm}{!}{\includegraphics[]{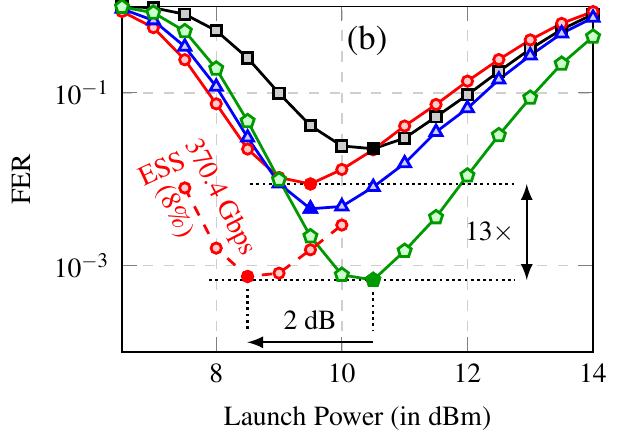}}
\resizebox{0.6\columnwidth}{!}{\includegraphics[]{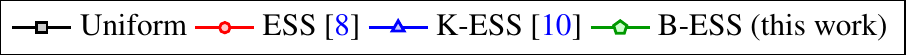}}
\vspace{-0.2cm}
\caption{{\bf Single-channel:} SNR (a) and FER (b) for transmission of 400 Gbps over 205 km fiber.}
\label{fig:SNR_DP_400G}
\end{figure} 

\vspace{-0.5cm}

In Fig.~\ref{fig:SNR_DP_400G} (a), effective SNR is shown for single-channel over 205 km fiber.
ESS has a 0.41 dB SNR penalty against uniform signaling, most of which is recovered by K-ESS.
As discussed in~\cite{gultekin_kess_arxiv}, this is because ESS has high kurtosis which K-ESS was designed to decrease.
B-ESS provides 0.33 dB SNR gain over uniform signaling thanks to its smaller energy variations.
We also see that the maximum SNR is achieved at a larger launch power for B-ESS (e.g., 1 dB larger than that for ESS), indicating that it is the most NL-tolerant scheme.
In Fig.~\ref{fig:SNR_DP_400G} (b), frame error rate (FER) is shown for $\nch=1$.
We see that while K-ESS brings limited gains, B-ESS provides 13 times decrease in FER against ESS.
To translate this FER improvement into data rate gain, we decreased the transmission rate of ESS-based signaling until we reached the same FER (dashed) as B-ESS.
This is achieved at 370.4 Gbps which means in this setup, B-ESS provides an 8\% data rate increase over ESS.
The reason why the 370.4 Gbps ESS has a smaller optimum launch power (e.g., 2 dB smaller than that of B-ESS) is that its kurtosis is larger than that of 400 Gbps ESS~\cite[Fig. 2]{gultekin_kess_arxiv}, and thus, it experiences an increased NLI.

\begin{figure}[h]
\centering
\resizebox{7cm}{!}{\includegraphics[]{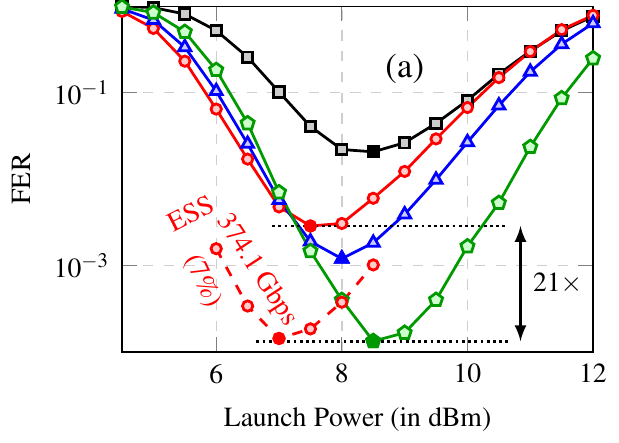}} 
\resizebox{7cm}{!}{\includegraphics[]{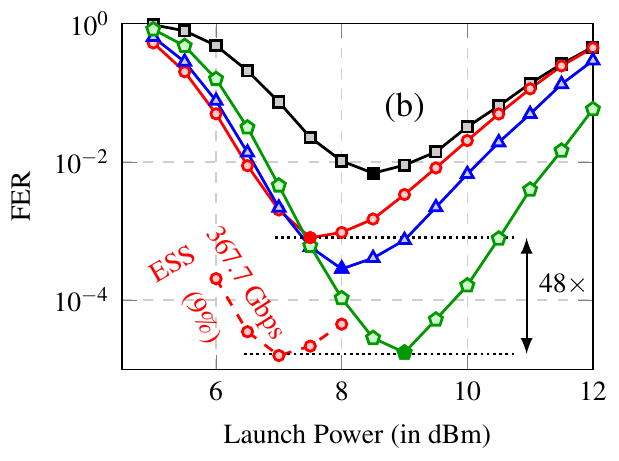}} 
\resizebox{0.6\columnwidth}{!}{\includegraphics[]{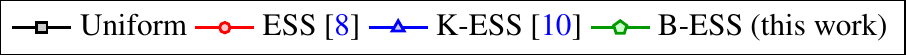}}
\vspace{-0.3cm}
\caption{{\bf WDM:} FER for transmission of 400 Gbps over 195 km with 55 (a) and 75 GHz (b) spacing.}\label{fig:SNR_DP_400G_wdm}
\end{figure} 

\vspace{-0.4cm}

In Figs.~\ref{fig:SNR_DP_400G_wdm} (a) and (b), FERs are shown for $\nch=5$ with $\bchsp=55$ and $75$ GHz, resp.
We see that B-ESS provides more than one order of magnitude decrease in FER in both cases.
When translated into data rate, the increase over ESS is 7\% for 55 GHz grid and 9\% for 75 GHz grid.
We intuitively attribute the gain for $\nch=5$ being smaller than that of $\nch=1$ when $\bchsp=55$ GHz, but being larger when $\bchsp=75$ to the balance between intra- and inter-channel NLI.
B-ESS effectively mitigates intra-channel NLI, leading to the gains observed for $\nch=1$.
Then for $\nch=5$, the strong inter-channel NLI experienced when the channels are placed very close to each other (i.e., $\bchsp=55$ GHz) somewhat degrades the performance, leading to a smaller gain.
However, B-ESS is able to mitigate (at least partially) inter-channel NLI when the channel spacing is relatively large, providing the largest gains as observed for $\nch=5$ with $\bchsp=75$ GHz.

\section{Conclusions}
We have proposed a novel shaping strategy to generate channel inputs with limited energy variations.
This is realized by considering a \underline{b}and-like portion of the enumerative sphere shaping (B-ESS) trellis.
Through end-to-end simulations, B-ESS is demonstrated to provide a considerable increase in data rates over uniform signaling and sphere shaping.
Future work includes optimization and implementation of B-ESS.

\vspace{2mm}

\footnotesize
{
\noindent \textbf{Acknowledgements:}~The work of Y. C. G\"{u}ltekin and A. Alvarado has received funding from the European Research Council (ERC) under the European Union's Horizon 2020 research and innovation programme (grant agreement No 757791).}



\end{document}